\documentclass[11pt]{article}

\usepackage{amssymb,amsmath,latexsym, hyperref, cite, natbib, graphicx, refstyle, pgfplotstable, booktabs, rotating, color, tabularx, authblk, array, url, soul, longtable,lscape,afterpage,cclicenses}

\usepackage{multirow}
\newcolumntype{C}[1]{>{\centering\arraybackslash}p{#1}}

\usepackage{geometry}
\usepackage{setspace}
\usepackage[multiple, hang,splitrule]{footmisc}


\usepackage[font=small, margin=0cm]{caption}

\begin{document}
\title{\textbf{Funding Data from Publication Acknowledgements: Coverage, Uses and Limitations}}
\author[1]{\textbf{Nicola Grassano}\thanks{n.grassano@sussex.ac.uk}}
\author[1]{\textbf{Daniele Rotolo}\thanks{Corresponding author: d.rotolo@sussex.ac.uk, @danielerotolo, Phone: +44 1273 872980}}
\author[1]{\textbf{Joshua Hutton}\thanks{j.r.hutton@sussex.ac.uk}}
\author[1]{\textbf{Fr\'{e}d\'{e}rique Lang}\thanks{f.lang@sussex.ac.uk}}
\author[1]{\textbf{Michael M. Hopkins}\thanks{m.mhopkins@sussex.ac.uk}}

\affil[1]{\small SPRU --- Science Policy Research Unit, University of Sussex, Brighton, United Kingdom}

\date{Version: \today \linebreak 
Accepted for publication in the \linebreak \textit{\textbf{Journal of the Association for Information Science and Technology}}\thanks{\href{http://dx.doi.org/10.1002/jasist.23737}{{\color{blue}DOI: 10.1002/jasist.23737}}. \copyright 2016 Grassano, et al.  Distributed under \href{http://creativecommons.org/licenses/by/4.0/}{{\color{blue}CC-BY-NC-ND}}.}}

\maketitle
\begin{abstract}
\onehalfspacing
\noindent This article contributes to the development of methods for analysing research funding systems by exploring the robustness and comparability of emerging approaches to generate funding landscapes useful for policy making. We use a novel dataset of manually extracted and coded data on the funding acknowledgements of 7,510 publications representing UK cancer research in the year 2011 and compare these `reference data' with funding data provided by Web of Science (WoS) and MEDLINE/PubMed. Findings show high \textit{recall} (about 93\%) of WoS funding data. By contrast, MEDLINE/PubMed data retrieved less than half of the UK cancer publications acknowledging at least one funder. Conversely, both databases have high \textit{precision} (+90\%): i.e. few cases of publications with no acknowledgement to funders are identified as having funding data. Nonetheless, funders acknowledged in UK cancer publications were not correctly listed by MEDLINE/PubMed and WoS in about 75\% and 32\% of the cases, respectively. `Reference data' on the UK cancer research funding system are then used as a case-study to demonstrate the utility of funding data for strategic intelligence applications (e.g.\ mapping of funding landscape, comparison of funders' research portfolios). 
  \newline\newline
{\bf Keywords:} funding data; acknowledgments; \textit{precision}; \textit{recall}; bibliometrics; scientometrics; cancer research; MEDLINE/PubMed; ISI Web of Science.\par
\end{abstract}
\clearpage 

\section{Introduction}
In the present climate of fiscal austerity, research funding is falling in a number of countries \citep{BIS2015}. Questions about the efficiency of funding mechanisms are therefore very much in the minds of policy makers and science policy analysts. Some advocate closer monitoring and measurement to optimise the `yield' from research \citep{Chalmers2014}. Others suggest funders should consider the types of research they support, ensuring they maintain diversity \citep{Wallace2015}. In this context, understanding the relationship between funding inputs and research outputs has important implications for the design and management of research at the level of grants, programmes, agencies' portfolios, and even at the national and international funding landscape level (e.g.\ where interdependencies and complementarities between funders could be identified, and co-ordination can improve efforts to address particular diseases).

Funders are investing in the development of data infrastructures capable of supporting such efforts as well as of increasing the transparency around funding decisions. These include the RePORTER database of the US National Institutes for Health (NIH), the Gateway to Research of the Research Councils UK (RCUK), and the OpenAIRE of the European Commission. Some funders have also combined efforts in order to compare their research portfolios and to track national trends in funding \citep{UKCRC2014}. However, considering the diversity of funders that populate funding systems, these efforts, led by major funders, do not include data on the activities of numerous smaller organisations that also support research.  

In order to consider the research being funded in a given field or country and to reflect on changes to funding policies, it is desirable to obtain a comprehensive overview of the funding landscape by tracing inputs and outputs from a wide range of funders and researchers that are active in a given area. One promising source of data for this purpose is the information included in the acknowledgement \textit{paratext} of scientific publications where authors commonly give thanks for support \citep[e.g.][]{Costas2012, Dawson1998, Rigby2011}. In particular, funding information in acknowledgements has the potential to establish a direct link between funding inputs and research outputs on a grand scale, without the need to gain direct access to data via individual funders or researchers. This paper explores a number of approaches to using funding data derived from in publication acknowledgements. In doing so, we address key questions about the coverage of these data and explore the utility of mapping funding landscapes using these data.

Making systematic use of the funding information included in the acknowledgment sections of publications is, however, difficult due to the lack of standardization in the structure and content of acknowledgements --- in contrast to references, for example. In order to support the use of acknowledgement \textit{paratext} some databases, such as ISI Web of Science (WoS) and MEDLINE/PubMed, provide ready-extracted and coded funding data. Yet, uncertainties remain about the coverage of such sources \citep{Koier2014, Rigby2011}. The motivation for this paper is to resolve some of the uncertainties of working with funding data, in the hope that improving methods and resources for the use of these data can help to improve research policy.

Our analysis focuses on 7,510 publications in cancer research published in the year 2011 and involving at least one author affiliated with a UK research host organisation \citep[see][]{CRUK2014}. Cancer represents by far the largest single disease area studied in the UK, attracting an estimated 20\% of funding invested by the government and major research charities \citep{UKCRC2014}. For each publication in the sample, we extracted and coded funding data from the acknowledgment sections and used these data for three purposes: (i) to devise heuristics that can provide grounds for future systematic collection of funding data from acknowledgements; (ii) to provide one of the first assessments, in terms of \textit{recall} and \textit{precision}, of the coverage of funding data available from MEDLINE/PubMed and WoS; (iii) to illustrate how bibliometric analysis of funding data taken from acknowledgment sections can be used to generate intelligence on funding systems to inform policy making in terms of most active funders, co-funding (i.e.\ joint occurrence of funders in publication acknowledgements), and funders' portfolio profiles of funded research.

The paper proceeds as follows: The next section positions the contribution of the present paper within the growing bibliometric literature on the use of acknowledgment sections of scientific publications; we then describe the methodological approach used to collect and code the data and introduce a number of procedural points of guidance on how to recognise and isolate funding information in the acknowledgment sections, as well as on how to codify this information; in the Results section, we examine the extent to which authors acknowledge funders in their publications; we compare, in terms of \textit{precision} and \textit{recall}, our data with those extracted by queries of the same set of publications from MEDLINE/PubMed and WoS; we then use the data to illustrate applications in funding portfolio and landscape analysis; we conclude the paper by discussing the main results of the approach we propose as well as the caveats and limitations associated with it.

\section{Literature review and emerging questions}

The breadth of information on scientific research activity that is included in the acknowledgment \textit{paratext} of scientific publications has attracted the attention of bibliometric researchers since the 1970s. Pioneering work by \cite{Mackintosh1972}, \cite{Cronin1991}, and \cite{McCain1991} focused on taxonomising the information contained in these sections. \cite{Mackintosh1972}, for example, analysed the scientific articles published in the \textit{American Sociological Review} from 1940 to 1962 and defined three types of acknowledgments, i.e.\ facilities, access to data, and help of individuals. Cronin's (\citeyear{Cronin1991}) taxonomy specifically included: (i) paymasters (i.e.\ funding), (ii) moral support (e.g.\ use of equipment, access to data), (iii) dogsbody (e.g.\ secretarial support, editorial guidance, data collection and analysis), (iv) technical (e.g.\ access to know-how, guidance on statistical procurers), (v) prime mover (e.g.\ inspiration provided by principal investigator, project directors, mentor), and (vi) trusted assessor (e.g.\ feedback, critical analysis and comments from peers).

Scholars have subsequently exploited publication acknowledgements for different bibliometric and evaluation purposes. These include the analysis of patterns of \textit{subauthorship} or \textit{peer interactive communication}, such as formal or informal assistance and acknowledgments to reviewers \citep[e.g.][]{Cronin2006, Cronin1992, Cronin2004, Cronin2003, Tiew2002}, as well as the cataloguing of the funding authors received to produce their publications \citep[e.g.][]{Harter1992, Lewison1995, Lewison1994}. The latter has received extensive attention because of the potential for linking research funding inputs with scientific outputs.

A considerable effort in this direction was made by \cite{Dawson1998} with the development of the Wellcome Trust's Research Output Database (ROD). The ROD included bibliographic information of about 214,000 UK biomedical publications between 1988 and 1995 and funding data extracted from the acknowledgement sections of these publications. This database enabled researchers to examine, for the first time and on a relatively large scale, the mix of funding sources in specific biomedical domains \citep{Lewison1998} and the extent to which funded and unfunded publications differ in terms of type of research \citep{Lewison1999} and impact as measured on the basis of citation-based indicators \citep{Lewison1998a, Lewison2001, MacLean1998}. These studies provided seminal evidence, for example, that publications acknowledging funders are likely to receive more citations than those not reporting acknowledgment to funders. Yet, this relationship was also found to be dependent on the research domain and type of citation-based indicator used \citep{Cronin1999, Lewison2003}.

These studies provided important insights on the extent to which authors acknowledged their funders. \Tabref{review} reports a summary of these and subsequent studies. It is worth noting that substantial variation in funding acknowledgment frequencies exists --- varying with research focus, journal, funder, and country. For example, the ROD database provides an overview of funding in several biomedical fields, with around 64\% of the publications acknowledging funders \cite{Dawson1998}. However, this figure masks considerable variation with publications. For example, in gastroenterology, arthritis and malaria research funders were acknowledged in about 54\%, 60\%, and 80\% of the cases, respectively \cite[e.g][]{Lewison1999, Lewison1998, MacLean1998}. More recently, \cite{Costas2012} found that more than 50\% of the publications in natural sciences (e.g.\ molecular biology and biochemistry, chemistry, physics and astronomy, and geosciences) reported funders in their acknowledgments, while publications in applied/clinical sciences (engineering, clinical medicine), mathematics, social sciences applied to medicine, psychology, psychiatry, and behavioral sciences reported funders in only 20-50\% of  cases. Publications in economics, social sciences, and humanities and arts reported funding in even fewer cases, typically less than 10\%. Similar patterns were found by \cite{Diaz-Faes2014} on a sample of about 40,000 publications published in 2010 by Spanish researchers. On the one hand these figures highlight differences in the funding required to undertake research (e.g.\ between humanities and natural sciences), but on the other hand differences may also reflect norms of reporting at different times and in different places.  

Acknowledgement patterns in journals might be expected to follow the rates of acknowledgements in the field they cover. For example, \cite{Zhao2010} showed that articles published in library and information science journals (e.g.\ \textit{Journal of the American Society for Information Science}, \textit{Information Processing and Management}, \textit{Journal of Documentation}, and \textit{College and Research Libraries}) include funding acknowledgements from about 6\% to 35\% of the cases (see \Tabref{review}), while Rigby's (\citeyear{Rigby2011, Rigby2013}) analysis of articles included in Cell revealed much higher funding acknowledgement rates (94\%). Great variations were found across journals in social sciences and humanities, such as psychology, sociology, history, and philosophy \cite{Cronin1993}: articles in \textit{Psychological Review}, \textit{American Sociological Review}, \textit{American Historical Review}, and \textit{Mind} acknowledged funders in 70.3\%, 53.5\%, 29.7\%, and 3.0\% of the cases, respectively.

\begin{spacing}{1}\scriptsize
\begin{longtable}{p{3.5cm}p{6.5cm}cc}
\caption{\label{tab:review}Reporting of funding data in publication acknowledgements in terms of journals, research domains, funders, and countries.} \\
\hline\hline
\textbf{Studies by main}&	\textbf{Data source}&			\textbf{Sample} &	 \textbf{\% of records reporting} \\
\textbf{analytical focus}&	\textbf{(period of observation)}&	\textbf{ } & 		\textbf{funding data} \\
\hline
\endfirsthead
\caption{Reporting of funding data in publication acknowledgements in terms of journals, research domains, funders, and countries\textit{(continued)}.} \\
\hline\hline
\textbf{Studies by main}&	\textbf{Data source}&			\textbf{Sample} &	 \textbf{\% of records reporting} \\
\textbf{analytical focus}&	\textbf{(period of observation)}&	\textbf{ } & 		\textbf{funding data} \\
\hline
\endhead
\\
\textbf{Journal(s)}                                              &                                                                       &         &               \\
\\
\cite{Cronin1991}                                        & \textit{J. of the Am. Soc. for Inf. Science} (1970-1990)                       			& 938     & 29.1\%       \\
\\
\cite{Harter1992}                                        & \textit{J. of the Am. Soc. for Inf. Science} (1972-1974, 1982-1984, 1988-1990)	& 391     & 32.2\%       \\
\\
\cite{Cronin1993}                                     	& Sample of journals (1971-1990):                                       &         &               \\
                                                                 	& \addtolength{\leftskip}{1em}	\textit{Psychological Review}                              			& 629     & 70.3\%       \\
                                                                 	& \addtolength{\leftskip}{1em}	\textit{Am. Sociological Review}                                               & 1,186    & 53.5\%       \\
                                                                 	& \addtolength{\leftskip}{1em}	\textit{Am. Historical Review}                                                 & 464     & 29.7\%       \\
                                                                 	& \addtolength{\leftskip}{1em}	\textit{Mind}                                                                  & 1,027    & 3.0\%        \\
\\
\cite{Cronin1999}                            		& Four information science journals (1989-1993)                         & 716     & 26.5\%       \\
\\
\cite{Tiew2002}                                           & \textit{J. of Natural Rubber Research} (1986-1997)                             & 310     & 20.3\%       \\
\\
\cite{Cronin2003,Cronin2004}          		& Sample of journals (1900-1990):                                       &         &               \\
        								& \addtolength{\leftskip}{1em}	\textit{Psychological Review}                                       & 2,707    & 31.9\%       \\
                                                                 	& \addtolength{\leftskip}{1em}	\textit{Mind}                                                                  & 1,850    & 3.4\%        \\
\\
\cite{Cronin2006}                                        & Cells (1975, 1985, 1995, 2004)                                        & 1,106    & 86.8\%-96.6\% \\
\\
\cite{Zhao2010}                                          & Sample of journals (1998):                                            &         &               \\
                                                                 	& \addtolength{\leftskip}{1em}	\textit{J. of the Am. Soc. for Inf. Science}                                   & 99      & 35.4\%       \\
                                                                 	& \addtolength{\leftskip}{1em}	\textit{Inf. Processing and Management}                                        & 46      & 34.8\%       \\
                                                                 	& \addtolength{\leftskip}{1em}	\textit{J. of Documentation}                                           & 24      & 33.3\%       \\
                                                                 	& \addtolength{\leftskip}{1em}	\textit{Library and Inf. Science Research}                                     & 17      & 29.4\%       \\
                                                                 	& \addtolength{\leftskip}{1em}	\textit{Library Trends}                                                        & 33      & 12.1\%       \\
                                                                 	& \addtolength{\leftskip}{1em}	\textit{Library Quarterly}                                                     & 12      & 16.7\%       \\
                                                                 	& \addtolength{\leftskip}{1em}	\textit{College and Research Libraries}                                   & 35      & 5.7\%        \\
\\
\cite{Rigby2011}                                                    & Sample of journals (-):                                               &         &               \\
                                                                 & \addtolength{\leftskip}{1em}	\textit{Cell}                                                                  & 301     & 94.0\%       \\
                                                                 & \addtolength{\leftskip}{1em}	\textit{Physical Review Letters}                                               & 3,414    & 83.0\%       \\
\\
\cite{Rigby2013}                                       & \textit{J. of Biological Chemistry} (2009)                                    & 3,596    & 89.9\%       \\
\\
\\
\textbf{Research domain(s)}                                               &                                                                       &         &               \\
\\
\cite{Lewison1995}                                             		& Sample of UK biomedical articles (1988-1992)                          		& $\sim$122,000 & 61.0\%       \\
\\
\cite{Dawson1998}                                              	& Wellcome Trust's Research Outputs Database (ROD) (1988-1995)                                		& 214,364  & 63.6\%       \\
\\
\cite{Lewison1998a, Lewison1998, Lewison2001}  	& Sample of gastroenterology articles (1988-1994)                      		& 12,925   & 54.4\%       \\
\\
\cite{MacLean1998}                                            	& Sample of malaria articles (1989)                                     			& 758     & 80.5\%       \\
\\
\cite{Lewison1999}                                         		& UK articles in arthritis research (1988-1995)                         			& 6,672    & 60.3\%       \\
\\
\cite{Lewison2003}                                        		& Articles in language therapy research (1991-2000)                     		& 1,048    & 51.0\%       \\
\\
\cite{Giles2004}                                         			& Articles in CiteSeer (1990-2004)                                      				& 335,000  & 56.1\%       \\
\\
\cite{Shapira2010a, Wang2011}                 		& Sample of nanotechnology articles (2008-2009)                         		& 91,164   & 66.9\%       \\
\\
\cite{Costas2012}                      		& WoS articles (2009)                                                   					& 1,253,909 & 43.2\%\\
\\
\hline\hline
\\
\\
\\
\textbf{Funder(s)}                   	&                                                                                        &        &         \\
\\
\cite{Lewison1994}               	& Sample of articles supported by the EU BAP as reported by surveyed authors (1986-1992) 	& 584    & 72.9\% \\
\\
\cite{Butler2001}                	& Articles funded by the Australian NHMRC (1994-1995)     		& 2,962   & 75.2\% \\
\\
\cite{Koier2014}     			& Articles funded by the Dutch climate programmes (2009-2012)                            			& 221    & 52.9\% \\
\\
\\
\textbf{Countries}                    	&                                                                                        	&        &         \\
\\
\cite{Salager-Meyer2009}  	& Sample of medical articles (2005-2007):                        	&        &         \\
                             			& \addtolength{\leftskip}{1em}	\textit{Spain}                                                                                  	& 50     & 4.0\%  \\
                             			& \addtolength{\leftskip}{1em}	\textit{United States}                                                                          	& 50     & 26.0\% \\
                             			& \addtolength{\leftskip}{1em}	\textit{Venezuela}                                                                              	& 50     & 72.0\% \\
\\
\cite{Wang2012a}           		& Articles from 10 countries (2009):                                                     &        &         \\
                            			& \addtolength{\leftskip}{1em}	\textit{United States}                                                                          	& 379,321 	& 44.1\% \\
                             			& \addtolength{\leftskip}{1em}	\textit{China}                                                                         	& 126,931 	& 70.3\% \\
                             			& \addtolength{\leftskip}{1em}	\textit{Germany}                                                                                	& 102,927 	& 40.8\% \\
                             			& \addtolength{\leftskip}{1em}	\textit{United Kingdom}                                                                         & 99,832  	& 42.6\% \\
                             			& \addtolength{\leftskip}{1em}	\textit{Japan}                                                                                  	& 87,582  		& 43.0\% \\
                             			& \addtolength{\leftskip}{1em}	\textit{France}                                                                                 	& 73,186  		& 38.1\% \\
                             			& \addtolength{\leftskip}{1em}	\textit{Italy}                                                                                  		& 62,609  		& 33.1\% \\
                             			& \addtolength{\leftskip}{1em}	\textit{Canada}                                                                                 	& 60,723  		& 49.1\% \\
                             			& \addtolength{\leftskip}{1em}	\textit{Spain}                                                                                  	& 48,245  		& 51.6\% \\
                             			& \addtolength{\leftskip}{1em}	\textit{Australia}                                                                              	& 18,590  		& 44.9\% \\
\\
\cite{Diaz-Faes2014} 		& Articles produced by Spanish researchers (2010)                    & 38,257  		& 72.6\% \\
\\
\cite{Gok2015}            		& Articles from 6 countries (2008-2011)                                   	& 242,406 & 	$\sim$44\%  \\
\\
\hline\hline
\multicolumn{4}{l}{\footnotesize \textit{Source: Authors' elaboration.}}\\
\\
\end{longtable}
\end{spacing}

Countries with very different funding systems also can be expected to have markedly different acknowledgment rates. \cite{Butler2001} found that researchers funded by the Australian National Health and Medical Research Council (NHMRC) acknowledge their funder in 75\% of papers. \cite{Costas2012} found that articles authored by Chinese researchers acknowledge funders in about 65\% of cases. Publications authored by researchers in South Korea, Taiwan, Sweden, Finland, Denmark, Spain, and Canada acknowledged funders in more than 50\% of the publications, while funding was acknowledged in less than 40\% of the publications of authors based in Turkey, Greece, Iran, Poland, India, Italy, and Israel. Similarly, \cite{Wang2012a} has shown in a study of 10 countries that 70\% of publications from China, but only 33\% of those from Italy, carry funding acknowledgements. 

Beyond variation in the occurrence of funding acknowledgments, the analysis of the relationship between funding and impact has been a key application of funding data in publication acknowledgments. Evidence of the positive correlation between the presence of funding acknowledgments in a publication and the citations the publication received were found by \cite{Zhao2010} on a sample of articles published in library and information science journals. \cite{Boyack2011} examined over 2.5 million publication records from 1980 to 2009 in MEDLINE/PubMed and found that publications acknowledging funding from the US Public Health Service (including the NIH) were cited twice as much as publications authored by US researchers acknowledging no funding source. However, as discussed above, the relationship between funding and impact varies across journals, research domains, and countries. For example, \cite{Rigby2011, Rigby2013} found that there is a weak positive relationship between the number of acknowledged funding sources and the citation impact of the publications included in \textit{Physical Review Letters} and the \textit{Journal of Biological Chemistry} and that this relationship becomes statistically non-significant in the case of publications included in \textit{Cell}. Furthermore, a recent study by \cite{Gok2015} provides evidence that the variety of the acknowledged funding sources (i.e.\ the number of distinct funding sources) is positively related to the total number of citations a publication receives and to the top percentile citations. Yet, the funding intensity (i.e.\ the number of unique funders acknowledged in a publication over the number of authors listed in the publication) was found to negatively affect citation impact. 

With important findings beginning to accumulate from the study of funding acknowledgements, the question arises about the `quality' of the underlying data. Early work on the manual coding of acknowledgments data pointed towards the difficulty of using these data, attributed to the lack of structure in unstandardized free-form formats. There are challenges in terms of finding acknowledgement data in the full-text as well as in extracting funding data from text in a consistent manner. Acknowledgements are not solely found in specific sections called `acknowledgements'. Some journals have separate sections on funding, or include acknowledgements at the beginning or the end of the documents or even in small print without a section indicator. Building samples of data has typically relied on intensive manual work both to retrieve publications, for which the electronic access was not available, and to identify and classify the text of the acknowledgment sections of these publications (e.g.\ identification and characterisation of funders by type and nationality). Only a few attempts were made to develop algorithms capable of extracting publication acknowledgments and the funding data included in these \citep[e.g.][]{Giles2004}.\footnote{\cite{Giles2004} developed an algorithm to extract acknowledgements (and funding data included in these) of about 335,000 research documents within the CiteSeer computer science archive. This algorithm searched for lines of text that included acknowledgement information achieving a level \textit{recall} of about 90\% --- i.e.\ about 10\% of acknowledgment data were not retrieved.} 

For this reason, research on the use of acknowledgment funding data for large-scale bibliometric analyses was limited until in more recent years a new wave of studies has emerged with much higher sample sizes  \citep[e.g.][]{Costas2012, Diaz-Faes2014, Gok2015, Shapira2010a}. These studies have mostly been facilitated by the increasing availability of ready-classified information. WoS, for example, has recently provided access to the funding text of acknowledgments and to the list of funders and grant codes mentioned in these --- this information is reported in the ``FX'' and ``FU'' fields, respectively --- for publications from August 2008. 

Nonetheless, we have limited knowledge on the \textit{recall} and \textit{precision} of acknowledgement funding data gathered through automated routines and provided by publication databases such as MEDLINE/PubMed and WoS. For example, on a sample of 117 publications supported by ``Dutch climate programmes'', \cite{Koier2014} found that WoS did not correctly report the list of funders in about 51\% of the cases. This therefore raises questions of the extent to which these limitations of WoS funding data also occurs in other domains and on how much additional benefit remains to be gained from manual coding of funding data in acknowledgments as compared to relying on data from major databases. A second problem that emerges when working with acknowledgement data concerns the standardisation and disambiguation of funders' names. Algorithms can support this process, yet ambiguity remains with regard to the selection of the level of analysis and on the treatment of funders that have changed their names over time \citep{Rigby2011}.

This paper aims to address some of these concerns and to contribute to the extant literature in three ways. Firstly, we develop a series of `heuristics' to guide the collection and coding of acknowledgment funding data, and provide some cautionary advice on their use. This feeds into ongoing debates over a greater standardisation of acknowledgment sections, which can facilitate guidelines for authors of publications, and for analysts extracting and coding of information there included. Secondly, we examine the \textit{recall} and \textit{precision} of MEDLINE/PubMed and WoS in retrieving publications that reported acknowledgements to funders as well as the accuracy of the list of funders provided by these databases as compared to a sample of manually collected funding data on UK cancer research. Finally, we provide further elaboration on the value of carefully gathered funding data for profiling funders' portfolios and national funding landscapes.

\section{Data and methods}
Our empirical analysis is focused on the UK cancer research domain, which is particularly suitable for the purpose of this paper for two main reasons. First, cancer research has been characterised by intense funding activity over the last few decades. This has involved a large variety of funders --- from industry to government and philanthropic organisations \citep{Eckhouse2008}. Second, cancer research falls within the broader domain of biomedical research where the rates of reporting funding data in the acknowledgement sections of publications are relatively high \citep{Dawson1998}.

The search set out to capture all publications in the domain of cancer research, involving at least one author from a UK research host organisation, and published in the year 2011. Delineating a broad topic such as cancer is, however, a complex task from a bibliometric perspective due to the breadth of the `cancer' field. We therefore preferred to rely on the Medical Subject Heading (MeSH) classification of MEDLINE/PubMed, an extensively developed and evolving vocabulary of medical terms \citep{Leydesdorff2012, Petersen2016}. The MeSH classification is composed of several thousands of descriptor terms, representing topics in medical research --- 26,404 descriptors composed the 2011 MeSH classification. These descriptors are organised in a tree-like structure and assigned, through a standardised indexing process performed by examiners at the US National Library of Medicine (NLM), to MEDLINE/PubMed publications in order to classify the content of these at different levels of specificity.\footnote{A descriptor may belong to more than one branch of the MeSH tree and may be complemented by qualifiers that further specify a publication's content in relation to the assigned descriptor. For more details on the classification see \url{www.nlm.nih.gov/pubs/factsheets/mesh.html}.}

We defined cancer research using the MESH descriptor ``Neoplasms'' (which captures any of its children descriptors, such as different types of cancer). The ``Neoplasms'' descriptor is formally defined as: ``\textit{New abnormal growth of tissue. Malignant neoplasms show a greater degree of anaplasia and have the properties of invasion and metastasis, compared to benign neoplasms}''. We queried MEDLINE/PubMed for 2011 data in early 2013, by using the following search string: ``Neoplasms''[MeSH Terms].\footnote{Retrieving publications related to cancer by using the MeSH descriptor ``Neoplasms'' inevitably misses some of those publications that might be deemed by researchers or funders to be associated with `neoplasms'. For example, a funder or researcher may consider studies of angiogenesis in healthy tissue to be relevant for understanding how tumours develop a blood supply, but if these papers are not considered to actually study cancerous tissue they may not be understood as within the study of neoplasms per se, and they will not be coded using the ``Neoplasms'' descriptor (or its children descriptors). The publications collected by the search employed in this study are therefore referred to as a `sample', reflecting the fact that the search is exhaustive within the neoplasms field (including cancerous and pre-cancerous growths, but also non-cancerous growths, which we accept as a minor limitation), but not comprehensive of neoplasms by other definitions.} This returned 115,101 documents published globally in that year.\footnote{We used the electronic date, i.e.\ the earliest date when a document is made publicly available.}

MEDLINE/PubMed data were matched with SCOPUS data to retrieve bibliographic data on all authors' affiliation addresses as MEDLINE/PubMed provides affiliation data only for the first listed author in the given publication. Records between the two databases were matched on the basis of the ``PubMed unique identifier'' (PMID) field of MEDLINE/PubMed that is also included in SCOPUS records. SCOPUS-MEDLINE/PubMed match was, however, obtained for 98.1\% of the records. The full match was not achieved because of time lag in publication indexing between MEDLINE/PubMed and SCOPUS (even though data collection began in early 2013, a full year after the close of the target year).

\begin{figure}
\includegraphics[height=7cm]{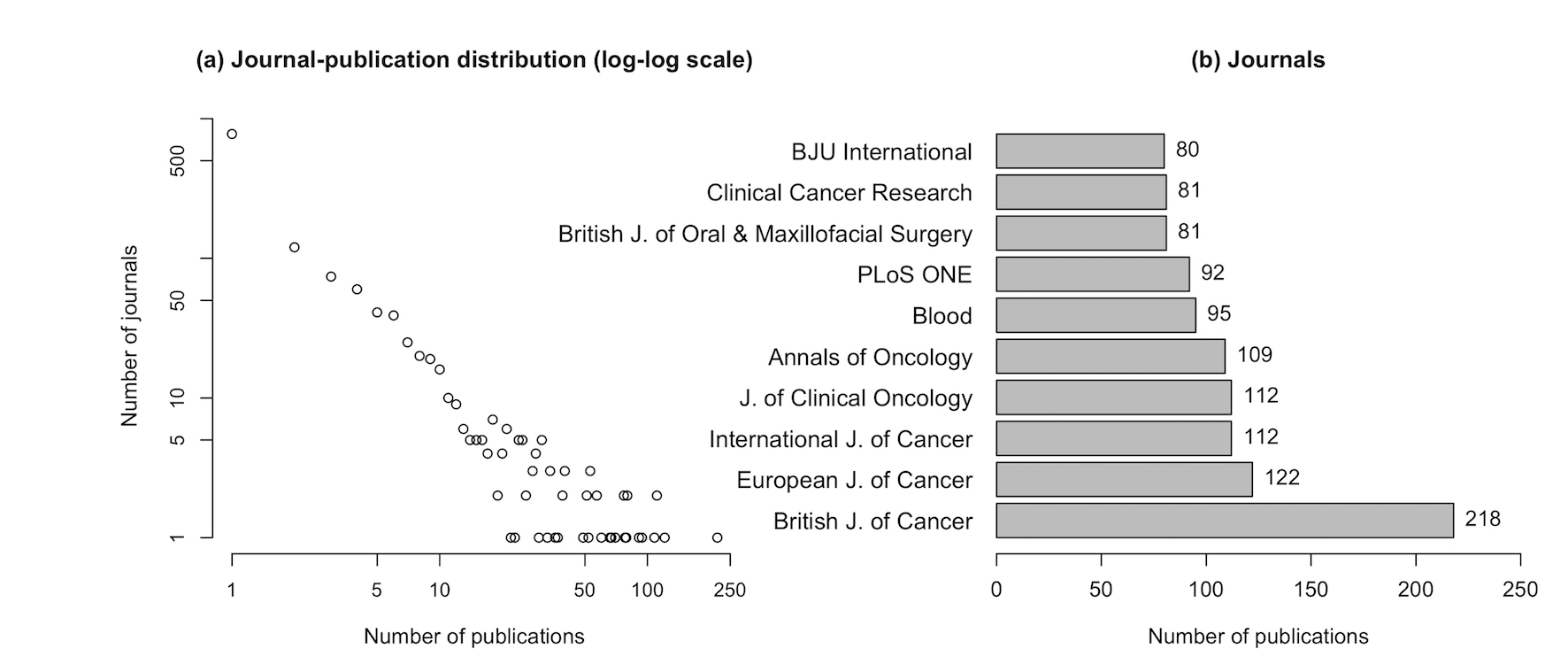}
\centering
\caption{Journal-publication distribution (a) and top-10 most frequent journals (b). \newline\textit{Source: Authors' elaboration.}}
\label{fig:journals}
\end{figure}

This match enabled us to identify UK publications as those involving at least one author based in a UK research host organisation --- unmatched records were manually screened.  We sought to include as many records as possible --- including all publication types (from editorials to reviews and all forms of research papers).

This process returned 7,922 publications, indicating that UK authors are represented in 6.9\% of the global production of publications on cancer for 2011. Publications were distributed across 1,449 journals. As depicted in \Figref{journals}a, the journal-publication distribution is skewed --- the \textit{British Journal of Cancer} is the journal in which UK publications on cancer were most frequently published in 2011 (see \Figref{journals}b). We were not able to electronically access 130 journals (712 records) due to access restrictions. E-mails to authors provided access for about about 42\% of publications included in these journals. As a result, we obtained access to the full-text of 7,510 publications (94.8\% of the initial sample of publications). For each of these, we collected and coded acknowledgements to funders that authors made, as detailed below.

\section{Extracting funding data}
The automatic extraction of funding information from acknowledgment sections in publications is challenging. As discussed above, authors use acknowledgments for a variety of purposes. Our focus is on financial support (e.g.\ external grants) that authors have received in order to undertake research that is then reported in a given publication. Funding data must therefore be separated from the text reporting other non-financial forms of support. For this reason, for each publication in our sample, we read the text of acknowledgments (including other sections that may report funding data), interpreted it, and extracted funding data manually. These data constitute the `reference data' on which we will assess the \textit{recall} and \textit{precision} of funding data provided by MEDLINE/PubMed and WoS. This section sets out steps taken to assure consistency in the data collection process, including the heuristics that were used to guide the coding of sample data (together with illustrative examples). These heuristics may be of assistance for future studies adopting similar approaches or aimed at developing routines capable of extracting funding information from acknowledgements in an automatic manner.

\setlength{\tabcolsep}{10pt}
\renewcommand{\arraystretch}{1.0}
\begin{table}\footnotesize
	\caption{\label{tab:cases}Cases of acknowledgment text and reporting of funding data (extracted funding information is represented with bold font).}
	\centering
{\begin{tabular}{cp{14cm}}
\hline\hline
\textbf{Case} & \textbf{Acknowledgement text}\\
\hline
1 & 	\textit{Acknowledgements} \newline 
	\textit{The authors thank our colleagues who have suggested or contributed data for current or previous versions of the database. They also would like to thank the swift and helpful advice from the ICGC DCC and BioMart team.}\newline
	
	\textit{Funding} \newline 
	\textit{\textbf{Cancer Research UK} (programme grant C355/A6253) and \textbf{FW6 EU project} MolDiag-Paca. R.C. is funded by \textbf{Breast Cancer Campaign}. Funding for open access charge: Cancer Research UK.}\newline
	\citep{Cutts2011}\\	
\\	
2 &	\textit{Acknowledgments}\newline
	\textit{We are grateful to Ms. Haruka Sawada and Ms. Noriko Ikawa for technical assistance. Our bio-repository is supported by funding from the \textbf{National Institute for Health Research} (UK) and the \textbf{Cambridge Biomedical Research Centre}. This work was supported by a Grant-in-Aid for Young Scientists (A) (22681030) from the \textbf{Japan Society for the Promotion of Science}.} \newline

	\textit{Disclosure Statement} \newline 
	\textit{The research was funded by  \textbf{OncoTherapy Science, Inc.} YI, YY, and KM are employees of OncoTherapy Science, Inc. YD, YN, and RH are scientific advisors of OncoTherapy Science, Inc.}\newline
	\citep{Takawa2011}\\
\\
3 &	\textit{Funding} \newline 
	\textit{\textbf{Grant Agency of Czech Republic} (303/09/0472 and 305/09/H008); \textbf{Ministry of Education of Czech Republic} (MSM0021620808 and 1M0505). Work at the Institute of Cancer Research is supported by \textbf{Cancer Research UK}.} Portions of this work were funded by \textbf{National Institutes of Health} (R01 ES014403 and P30 ES006096 to D.W.N. and Z.S.).\newline 

	\textit{Acknowledgments}\newline 
	\textit{The authors would like to thank the Grant Agency of the Czech Republic and the Ministry of Education of the Czech Republic.}\newline 
	\citep{Stiborova2012}\\
\\
4 &	\textit{Acknowledgments}\newline
	\textit{The authors wish to thank colleagues in the Histology Laboratory and the Electron Microscopy Unit, Veterinary Laboratory Services, School of Veterinary Science, University of Liverpool, for excellent technical support. T. Soare was supported by a scholarship from the Agency for Credits and Studies, \textbf{Romanian Ministry of Education, Research and Innovation}.}\newline
	\citep{Soare2012}\\
\\
5 &	\textit{Acknowledgments}\newline
	\textit{MG was supported by the grant Gu 1170/1-1 of the \textbf{German Research Foundation}. The support for using the research version of the Pinnacle treatment planning software from Philips Radiation Oncology Systems, Fitchburg, WI, USA is acknowledged. This work was partially supported by research Grant C46/A2131 from \textbf{Cancer Research UK}. We acknowledge \textbf{NIHR} funding to the NHS Biomedical Research Centre. The fruitful discussions and thoughtful proofreading of the manuscript by Kevin Brown and Joel Goldwein (Elekta) is acknowledged.}\newline
	\citep{Guckenberger2011}\\
\\
6 & 	\textit{Acknowledgments}\newline
	\textit{The study was supported by the grant MA1659/6-1/2 of the \textbf{Deutsche Forschungsgemeinschaft (DFG)}. The recombinant topoisomerase II was a generous gift of Fritz Boege, Institute of Clinical Chemistry and Laboratory Diagnostics, Heinrich Heine University (Duesseldorf, Germany). We thank Dr. Antonella Riva and Dr. Paolo Morazzoni (Indena SpA, Milan, Italy) for provision of test material. The authors have declared no conflict of interest.}\newline
	\citep{Esselen2011}\\

\hline\hline
\multicolumn{2}{p{14cm}}{\footnotesize \textit{Notes: Publications for cases 3 and 4 were electronically published in 2011, but included in a journal volume and issue in 2012.\newline
Source: Authors' elaboration.}}
\end{tabular}
}
\end{table}

The first step of data collection was the identification of the portions of publication full-text where funding data are reported. These data, as expected, are often included in the acknowledgment sections, which are usually positioned before or after references, or, in a few cases, at the beginning or in the footnotes of the publication. However, there are cases where funding information is reported in dedicated sections called, for example, ``Financial Support'', ``Financial Information'', or ``Funding''. These are often located next to the acknowledgment sections. Case 1 in \Tabref{cases}, for example, presented two separate sub-sections: ``Acknowledgements'' and ``Funding''. There are also cases of publications where funding information is reported in multiple sections. For example, in Case 2, the analysis of the acknowledgments section only would have inevitably missed funding to this publication coming from the industry, as accounted for in the disclosure of conflict section. 

Case 2 also provides indication of the importance of interpreting the wording used by authors to declare financial support. For example, it is crucial to distinguish between past and current financial support made available to the authors (e.g.\ historic honoraria as opposed to direct funding for research leading to a given publication). In addition, Case 2 reveals that the authors acknowledged their employer as funding the study. However, in many cases authors do not do so, and where they do not, it is important to note that we do not infer that their employer was the funder of the research. We have only extracted funding information that is explicitly declared by authors.\footnote{Previous studies on the use of funding data reported in publication acknowledgments have adopted two different coding approaches. 8 of the 30 studies listed in \Tabref{review} (notably those by Lewison and colleagues) have complemented the information on funding reported by authors in acknowledgments with the information on authors' affiliations (e.g.\ when an author of a given publication was found to be affiliated to a firm, the firm was included in the list of funders of the publication). The remaining studies in \Tabref{review} (22 out of 30) have instead exclusively focused on the funding information included in the acknowledgments. Our study adopts the latter approach. This, in turn, enables us to make a broader comparison with the results of previous studies as well as to conduct a more precise comparative analysis on \textit{recall}and  \textit{precision} of WoS data, which are based on the text of publication acknowledgments only.}

Once the relevant funding information for a given publication is identified, the information included in this section requires interpretation before being coded, i.e.\ before the list of funders can be defined. In certain cases, identifying the list of organisations that financially supported a publication is relatively straightforward. There are, however, instances where this distinction is ambiguous because of the general usage of the term `support' in the English language. As discussed, authors often use this term to give credits to a variety of types of support such as financial, material, or moral \citep[e.g.][]{Cronin1991}. For this reasons, we included an organisation in the list of funders if we could infer, with a certain degree of confidence, from the acknowledgment text that the `support' from a given acknowledged organisation was financial. For example, in the Case 3, we interpreted that  \textit{``supported by Cancer Research UK''} referred to financial support because this funder was acknowledged in a dedicated funding section. In those cases where there was no clear way to detect the financial nature of the support acknowledged, we adopted a conservative approach and assumed that the support was not financial --- the acknowledged organisation was therefore not included in the list of funders.\footnote{`Travel support', `in-kind donations' and some sorts of support (technical, lab access, access to journals and the like) are often acknowledged by authors. To choose what constitute acknowledgment for funding and what does not, we follow our principle to codify as funding support only when we conclude there is a direct link between the publication and monetary support. So of the three items mentioned, only travel support has to be codified, given some money were given to pay for travel during the course of the research.}

After identifying the relevant funding information, the last step was to codify this information in a consistent manner. To do so, we first extracted the full name of an organisation as stated in the acknowledgment text and added the ISO 2-digit code of the country where the organisation is located as we could infer from the acknowledgment text or authors' addresses. Funders that were acknowledged more than once in the same publication were counted only once. Harmonisation of names was undertaken using the highest organisation level of a funder in most cases. For example, in Case 4, the highest organisational level is the ministerial level, i.e.\  \textit{Ministry of Education, Research and Innovation (RO)}.

The language in which the name of the funder is reported can also be an issue. A funder based in a non-English speaking country can have its name reported in the acknowledgement either in the native language (so expressed in a language other than English) or in its translated English form. The  acknowledgments of Case 5 and 6 in \Tabref{cases} are a clear example of this, given that the  \textit{German Research Foundation} and the  \textit{Deutsche Forschungsgemeinschaft} are the same organisation.

Finally, also the task of adding next to each funder name the ISO 2-digit code of the country where it is located was not unambiguous. A funder's country of origin is not always clearly identifiable by its name or via a web search. In ambiguous cases, we assumed that the organisation acknowledged was from the same country as the author who acknowledged it if no further information was available. In case of multinational companies, we used the location of their headquarters as the home country.

Using the above heuristics helps to ensure consistency in the data collection and codification, but some degree of ambiguity persists, especially in cases where the wording of the acknowledgment was not clear. Consistent coding was ensured through use of a protocol containing illustrative examples (as above) and overseeing the coding process through periodic cross-checking of entries among the coding team.

\section{Results}
The sections below report our findings in relation to the extent to which authors acknowledge their funders in the field of cancer research, and on the \textit{precision} and \textit{recall} of MEDLINE/PubMed and WoS in retrieving UK cancer publications that report funding information in their acknowledgments. We also examine the accuracy of the list of funders (acknowledged in publications) as provided by these databases. We then use the collected data to explore the UK cancer research funding system and to provide examples of the strategic intelligence that such data can provide to analysts and policy makers.

\subsection{Do UK authors of cancer research acknowledge their funders?}
The analysis revealed that 52.1\% (3,914 out 7,510 publications) of the sample disclosed at least one funder.\footnote{The percentage of publications reporting funding data increases to 57.0\% (3,741 out 6,560 publications) when considering only articles and reviews as classified by SCOPUS.} We further examined our sample of publications below to explore whether authors might not be reporting funding information. We first distinguished publications in our sample in three sub-samples: (i) publications with funding data in acknowledgments (52.1\%), (ii) publications with no funding data in acknowledgments (30.4\%), and (iii) publications with no acknowledgement sections (17.4\%).

\begin{figure}[h]
\includegraphics[width=\textwidth]{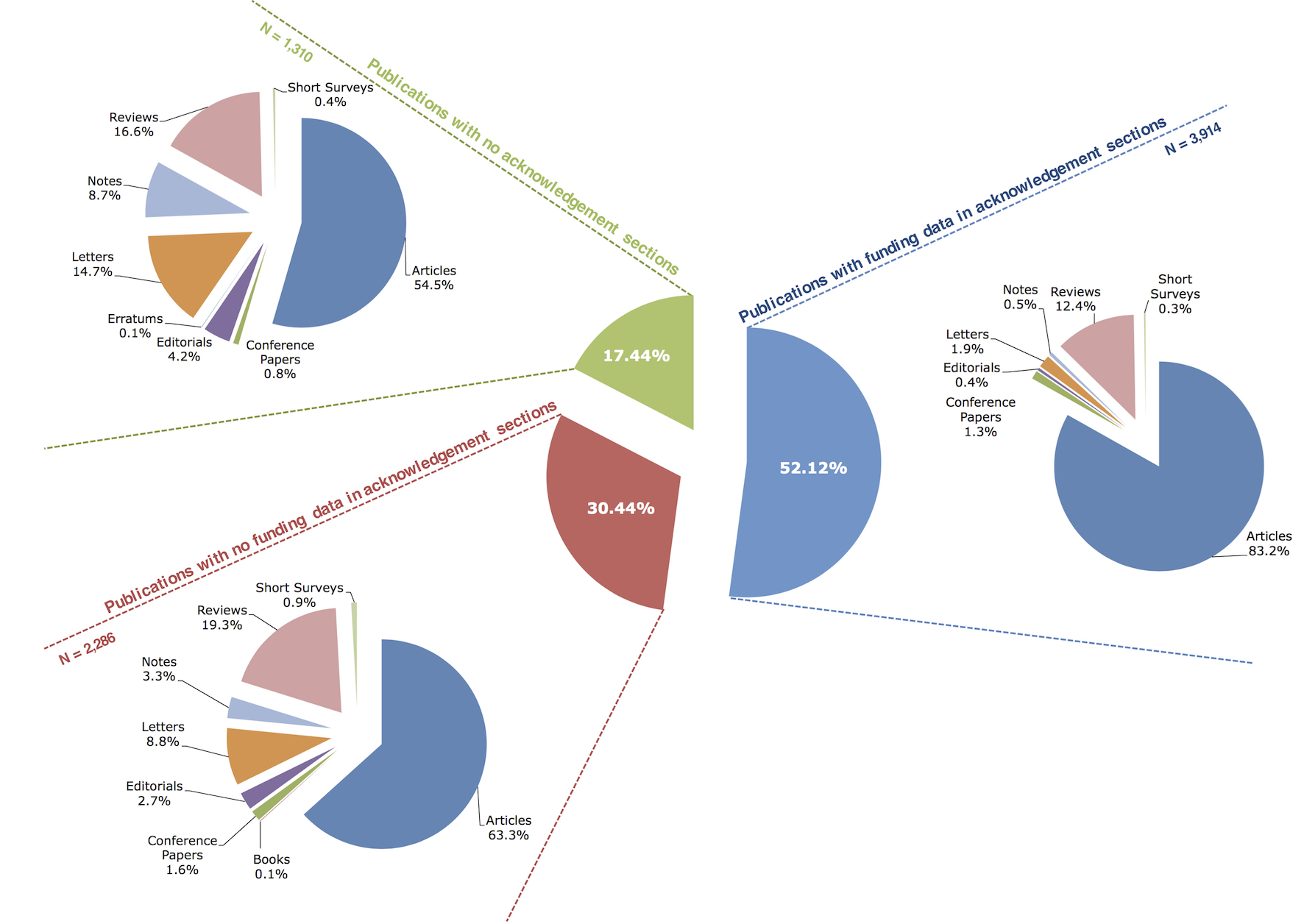}
\centering
\caption{Reporting of funding data by type of publication.
\newline\textit{Source: Authors' elaboration of the basis of MEDLINE/PubMed and SCOPUS data.}}
\label{fig:coverage}
\end{figure}

\begin{figure}
\includegraphics[height=7cm]{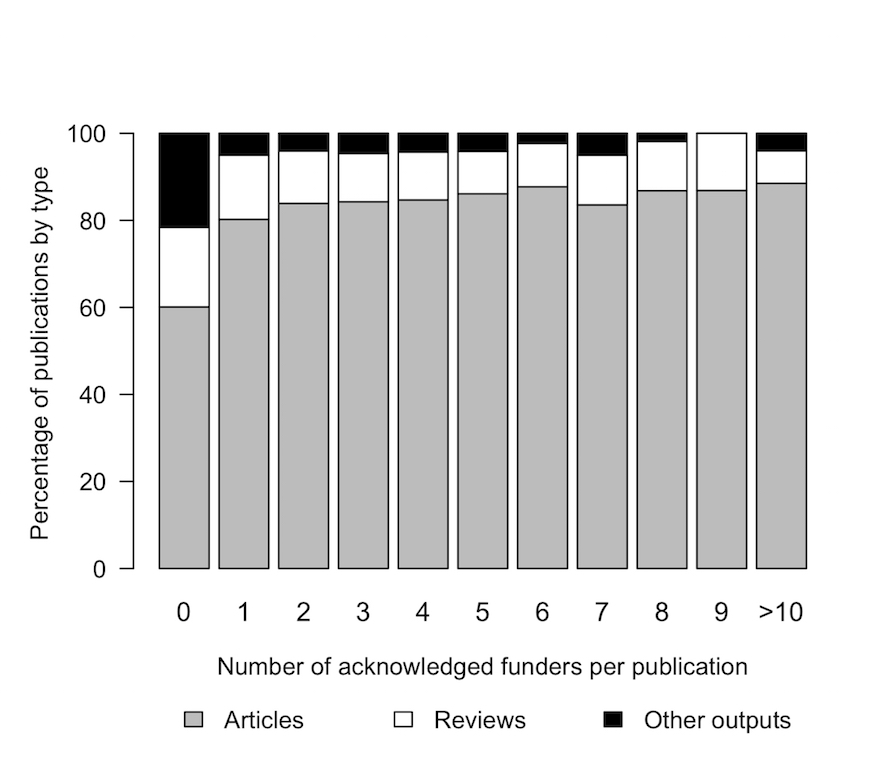}
\centering
\caption{Number of acknowledged funders by type of publication (``Other outputs'' include: books, conference papers, editorials, errata, letters, notes, short surveys).
\newline\textit{Source: Authors' elaboration of the basis of MEDLINE/PubMed and SCOPUS data.}}
\label{fig:publication_types}
\end{figure}

Publications in each of these sub-samples were then classified by type on the basis of the classification of publication records provided by SCOPUS, i.e.\ articles, reviews, conference papers, editorials, errata, notes, letters, and short surveys. The results of this analysis, which are depicted in \Figref{coverage}, provide evidence that the sub-samples of publications with no acknowledgment sections and with acknowledgements but no named funders are composed to a much greater extent of `less cost-intensive' publications (conference papers, editorials, errata, notes, letters, and short surveys) than within the sub-sample of publications with acknowledgements, i.e. 45.5\% and 36.7\% against 16.8\%. These findings are also in line with previous research \citep[e.g.][]{Salager-Meyer2009}. Also, about 22\% of the publications with no acknowledgment to funders are represented by `less cost-intensive' research outputs, while this proportion significantly reduces --- it ranges from 0\% to 5\% --- when considering publications with acknowledgment to at least one funder (see \Figref{publication_types}). 

We then explored two additional explanations for the lack of named funders in publication acknowledgments: (i) the publication required research funding, but relevant information was omitted from the publication full-text either by the authors or the publisher; or (ii) the publication was supported by the authors' employers, which are indirectly acknowledged through the authors' affiliations. To determine how these works were funded, we examined the two sub-samples of publications with no funding data as follows. First, we randomly selected more than 10\% (208) of the publications within the sub-sample of publications with no acknowledgment sections. A closer examination of this sub-sample revealed that about 38\% (79/208) of publications could be classified as case reports. These are likely to require no external funding other than the support of authors' employers. We therefore did not further examine this category of outputs. For the remaining publications (129/208) an e-mail query was sent to the corresponding authors. We obtained a response rate of about 38\% (49 replies). In about 18\% of responses, a funding contribution to the publication other than the authors' employers was revealed. In 61\% of responses, authors instead revealed a contribution by their employer, while in the remaining 21\% of the cases, authors indicated that the publications did not require any funding. In summary, we can estimate that 11.2\% (0.18*0.62) of the publications with no acknowledgment sections were actually supported by an external funder, i.e.\ less than 2\% (0.11*1310/7510) of the entire sample of publications.

We then performed a similar analysis for the sub-sample of publications with acknowledgements but no stated funders. 69\% of these publications explicitly stated that they did not benefit from financial support. Of the remaining 31\% of publications, we randomly selected a sub-sample of more than 10\% (89) publications for further investigation. On reading the papers we found that about 16\% (14/89) of the publications could be deemed to be case reports and therefore it was assumed no funding (beyond the authors' employers) was necessary to produce the publication. Financial support for the remaining publications (75) was queried via e-mail to the corresponding author, and a response rate of 37\% (28 responses) was obtained. About 14\% of responders revealed contribution by funders other than authors' employers; 50\% of the responses suggested that the research was supported by their employers; authors indicated the remaining publications did not require any funding. We can therefore estimate that 3.6\% (0.31*0.84*0.14) of the publications with acknowledgements but no stated funders were externally funded, i.e. about 1\% (0.04*2286/7510) of the entire sample of publications.

On the basis of the findings above, we concluded that when a UK cancer publication reports no funding data, it is likely that the publication was `less cost-intensive' (e.g.\ editorials, notes, letters, case reports) and therefore did not require any additional funding or that the publication was supported by authors' employers. From a review of the text of acknowledgements and survey of a sample of the authors, it seems that the omission of all funding information is unlikely --- we estimate overall less than 3\% of the entire sample. However, it is clear that sometimes authors do not think it is relevant to acknowledge their employers for funding the publication --- the disclosure of their address somewhat fulfils this purpose.

\subsection{\textit{Precision} and \textit{recall} of MEDLINE/PubMed and WoS}
To enable the comparison of the funding data extracted through the process described above with those available from MEDLINE/PubMed and WoS, we first harmonised the names of the funders listed in the acknowledgements of the publications included in our sample.\footnote{The Vantage Point software package aided the harmonisation of organisation names} This led to the identification of 2,549 distinct funding organisations of which 663 organisations in the public or charitable sector were based in the UK. A further 1,579 public or charitable sector organisations were identified outside the UK. Industry (e.g. pharmaceutical companies) were classified separately, with 307 firms being identified. About 64\% of the publications reporting funding information recognised support from two or more funders, with authors of cancer research publications acknowledging an average of 3.3 funders per publication. We compared these data, namely the `reference data', with those available in MEDLINE/PubMed and WoS across few descriptive indicators and in terms of  \textit{precision} and  \textit{recall}.\footnote{For the sake of clarity, we defined \textit{recall} as the ratio between \textit{true positives} (i.e.\ the number of publications for which both MEDLINE/PubMed or WoS and the `reference data' reported at least one funder) and the sum of \textit{true positives} and \textit{false negatives} (i.e.\ the number of publications for which these databases did not report funding data, but we found funding information in publications' full-text). \textit{Precision} was instead defined as the ratio between \textit{true positives} and the sum of \textit{true positives} and \textit{false positives} (i.e.\ the number of publications for which MEDLINE/PubMed or WoS reported funding data, but our analysis revealed no funding information in publications' full-text).} This comparison is reported in \Tabref{comparison}.

\setlength{\tabcolsep}{7pt}
\renewcommand{\arraystretch}{1.0}
\begin{table}\footnotesize
	\caption{\label{tab:comparison}\textit{Recall} and \textit{precision} of MEDLINE/PubMed and WoS funding data.}
	\centering
{\begin{tabular}{lcccc}
\hline\hline
	& \multicolumn{4}{c}{\textbf{Database}}\\
	\cline{2-5}
	&	\textbf{Reference data}&	\textbf{MEDLINE/PubMed}& \multicolumn{2}{c}{\textbf{ISI Wed of Science (WoS)}}\\
\hline
Number of publications &				7,510&		7,510&		\multicolumn{2}{c}{7,082}\\
\\
Publications  reporting&		3,914&		1,712&		\multicolumn{2}{c}{3,736}\\
funding data						&		(52.1\%)&		(22.8\%)&		\multicolumn{2}{c}{(52.7\%)}\\
\\
\textit{Recall}&						-&			41.9\%&		\multicolumn{2}{c}{92.8\%}\\
\\
\textit{Precision}&					-&			95.7\%&		\multicolumn{2}{c}{94.3\%}\\
\\
			&					&				&		\textit{Before}& \textit{After}\\
			&					&				&		\textit{harmonization}& \textit{harmonization}\\
Number of distinct& 			2,549& 		17& 			6,714& 						3,541\\
funders\\
\\
Number of funders\\
per publication \\
\addtolength{\leftskip}{1em}	\textit{Mean}&  						3.3& 			1.4 & 		4.1& 							3.7\\
\addtolength{\leftskip}{1em}	\textit{Std. Dev.} & 					(5.0)& 		(0.8)& 		(6.1)&  						(5.3)\\
\addtolength{\leftskip}{1em}	\textit{Max} & 						78    & 		6       & 		92    & 						76      \\
\\
\textit{Accuracy} of the list    & 		-     & 		24.8\%& 		-     & 						68.0\%\\
 of funders\\
\hline\hline
\multicolumn{5}{l}{\footnotesize \textit{Source: Authors' elaboration.}}
\end{tabular}
}
\end{table}

MEDLINE/PubMed reported funding information only for 23\% of publications (1,712 records). These publications overlapped with those in the `reference data' in about 96\% of the case (1,639 records), while the remaining 4\% (73 records) reported funding data even though we could not identify funding sources in the publications' full-text (\textit{false positives}). In other words, MEDLINE/PubMed did not report funding data for 2,275 publications (\textit{false negatives}). Accordingly, MEDLINE/PubMed funding data have very low \textit{recall}, i.e. 41.9\% (1639/(1639+2275)) and relatively high \textit{precision} of 95.7\% (1639/(1639+73)). Yet, MEDLINE/PubMed publications with funding data identified only 17 distinct funders and 1.4 funders per publication against the 2,549 funders and the 3.3 funders per publication reported in the `reference data'. This reflects the focus of MEDLINE/PubMed on major US funders and few large non-US funding organisations.\footnote{See \url{www.nlm.nih.gov/bsd/grant_acronym.html}} As a result, funders are not correctly listed in 75.2\% of the 1,639 \textit{true positives}. About 73\% of these cases miss at least one funder, while less than 2\% of them report at least one funder not acknowledged in publications' full-text.

In the case of WoS, we first matched our data with WoS data.\footnote{The matching was performed by using the \textit{medlineR} routine \cite{Rotolo2015a}.} 7,082 out of 7,510 publications (about 94.3\%) included in the `reference data' were also found in WoS. We then harmonised funders' names listed in the ``FU'' field of WoS data in order to make WoS data comparable with the `reference data'. The proportion of publications reporting funding data is 52.7\% (3,736 records), i.e.\ similar in proportion to the `reference data'. 3,524 publications of these are also included in the `reference data' (\textit{true positives}), while WoS did not report funding data for 274 publications for which the `reference data' included funding information (\textit{false negatives}). 212 publications with no funding information in the `reference data' were instead found to contain funding data in WoS (\textit{false positives}). This indicates that within the UK cancer research domain WoS data \textit{recall} is of 92.8\% (3524/(3524+274)) and its \textit{precision} is 94.3\% (3524/(3524+212)). These findings also suggest that the \textit{recall} and  \textit{precision} of WoS funding data vary across research domains --- \cite{Koier2014} found that WoS did not correctly recognise and retrieve acknowledgement sections in about 24\% of cases of research publications sponsored by the ``Dutch climate programmes'' in the field of climate change research.

The number of distinct funders listed in WoS data (after the harmonisation of their names) is much higher than those identified in the `reference data': 3,541 against 2,549. This is also evident from the average number of funders per publications, i.e.\ 3.7 in WoS data and 3.3 in the `reference data'. To further investigate these differences, we focused on the sample of publications for which both the `reference data' and WoS reported funding information, i.e. 3,524 publications. Within this sample, WoS did not report the same number of funders as in the `reference data' in 32.0\% of the cases: in 10.5\% of the records WoS missed at least one funder, while at least one funder more than in the `reference data' was included by WoS in 21.5\% of the sample. This includes cases where organisations were acknowledged, but under the guidance provided here (in Section 4) would not be classified as providing financial support. For example, the acknowledgments of a publication in the sample stated:  

\bigskip

``\textit{[...] \textbf{This study was not funded}. GPRD operates within the MHRA. GPRD has received funding from the MHRA, Wellcome Trust, Medical Research Council, NIHR Health Technology Assessment programme, Innovative Medicine Initiative, UK Department of Health, Technology Strategy Board, Seventh Framework Programme EU, various universities, contract research organisations and pharmaceutical companies. The Department of Pharmacoepidemiology and Pharmacotherapy, Utrecht Institute for Pharmaceutical Sciences, has received unrestricted funding for pharmacoepidemiological research from GlaxoSmithKline, Novo Nordisk, the private-public funded Top Institute Pharma (www. tipharma.nl, includes co-funding from universities, government, and industry), the Dutch Medicines Evaluation Board, and the Dutch Ministry of Health [...]''}  \citep{vanStaa2012}.\footnote{The publication was electronically published in 2011.}

\bigskip

Despite the authors? explicit declaration that the focal study did not receive any external funding, WoS data listed all the organisations, subsequently mentioned in the acknowledgments, as funders of the publication. This suggests that the ``FU'' field of WoS requires careful cleaning to remove cases where organisations are reported as funders despite the clear indication that they are listed as a declaration of potential conflict of interests (e.g.\, in the case of authors paid by companies for unrelated work) and for previous funding, rather than funding of the work contained in that particular paper.

In summary, the comparison provides evidence that in a very substantial number of cases, manual coding of the data reveals a different outcome to that provided by MEDLINE/ PubMed and WoS.

\subsection{Exploring UK cancer funding}
This section demonstrates how carefully gathered funding acknowledgements data, although labour intensive to prepare, can be used to provide research portfolio profiles for individual organisations as well as providing landscape overviews of wider funding environments. This is an important approach because alternatives based on self-reporting of funding inputs by research funders have a number of limitations including, \textit{inter alia}, reliance on funders to provide data and different data collection and coding conventions amongst funders \citep{Hopkins2013}. Although efforts are underway to bring together and classify funding inputs data (e.g.\ the UK Clinical Research Collaboration's coverage of 60 funders' self-reporting of funding allocations --- mentioned in the introduction) as the data explored in this section show, biomedical research funding involves a very large number of national and international funders, including governments, charities and firms, and funding landscape methodologies need to be able to take this diversity into account.

\subsubsection{The contribution of different types of funders}
\begin{figure}[h]
\includegraphics[height=14cm]{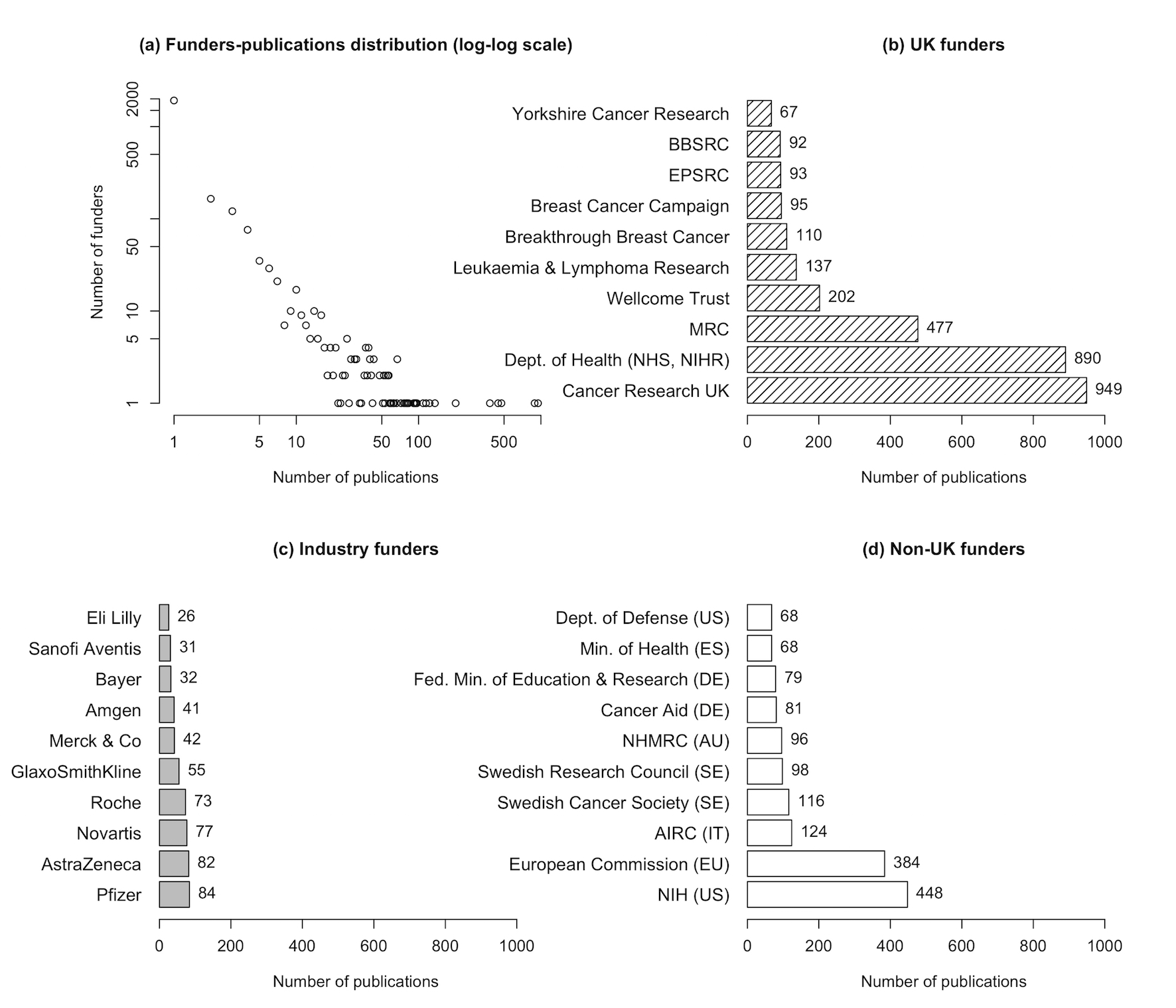}
\centering
\caption{Funder-publication distribution (a) and top-10 UK (b), industry (c) and non-UK (d) acknowledged funders.
\newline\textit{Source: Authors' elaboration.}}
\label{fig:funders}
\end{figure}

An advantage of collecting funding data from publications is the breadth of funders that it is possible to reveal without a priori knowledge of the field or privileged access to funders' data. As noted above, 2,549 funders were associated with at least one publication by a UK author in 2011, although, as reported in \Figref{funders}a, the distribution funder-publication is highly skewed. 75\% (1,921) of funders were associated with just a single publication in the sample. More than 296 public and charitable sector funders supported two or more publications. If we define `major UK funders' as those acknowledged in at least 2\% of the publications acknowledging funding in our sample, these supported about 50\% (1978/3914) of the publications. A further 37\% (1440/3914) of publications acknowledging funding were supported by `minor UK funders' (that is, funders that supported less than 2\% of funded publications individually). About 22\% (849/3914) of the sample acknowledged at least one major and one minor UK funder. Although clearly a relatively small number of funders fund the majority of UK cancer research, these descriptive statistics highlight the important role that the myriad of small funders play in the overall funding system.

\Figref{funders} reports the top-10 most acknowledged funders in the categories of UK public sector and charitable funders (\Figref{funders}b), industry (\Figref{funders}c), and non-UK public sector and charitable funders (\Figref{funders}d). The analysis suggests that international funders make a strong contribution to UK scientific output. The US National Institutes of Health (NIH) was, for example, acknowledged in 11.5\% of publications in the sample, i.e.\ in a similar proportion of publications that acknowledged UK Medical Research Council (MRC). The nature of this contribution is, however, more likely to be via support to the foreign collaborators of UK authors than to UK authors directly --- with unstructured acknowledgements it is impossible to tell which authors are supported by which funders most of the time.  The European Commission (EC) funding (acknowledged in 9.8\% of publications in the sample) is more likely to be provided as direct support to UK researchers, but much of this will also be linked to international collaborators. Nonetheless, given that 43\% of the sample of 7,510 publications involved an international collaboration, it is clear that the UK cancer research output is substantially supported by funding coming from countries like the US, Italy, Sweden and France (and the relationship goes both ways, clearly). Specifically, about 47\% (1853/3914) of the publications acknowledging funders mention at least a non-UK public sector or charitable funder. Of the 307 firms acknowledged for their funding in UK cancer publications, large pharmaceutical companies are the industrial actors most acknowledged in publications, led by Pfizer, AstraZeneca, and Novartis. Collectively these firms supported about 18\% (699/3914) of the publications acknowledging funders. About 20\% (141 publications) of these acknowledged at least two industrial actors for funding, thus providing some indication on the extent to which firms jointly support pre-competitive research. 

Funding acknowledgements also enable the exploration of the extent to which funders are jointly acknowledged in publications (co-funding). Use of the term co-funding for publications does not imply funders were aware of collaborative work \textit{ex-ante}. Rather, it is anticipated that funded researchers work together in ad-hoc combinations in order to drive their research forward. In some cases, co-funding will be as a result of individual researchers accessing multiple sources of funding in order to undertake their work. However, again, due to the unstructured nature of funding acknowledgments it is often difficult to tell how particular funders contributed to a given publication. Given that our data are focused on the UK context, we only examined the co-occurrence of UK funders in funding acknowledgements.

\begin{figure}
\includegraphics[width=\textwidth]{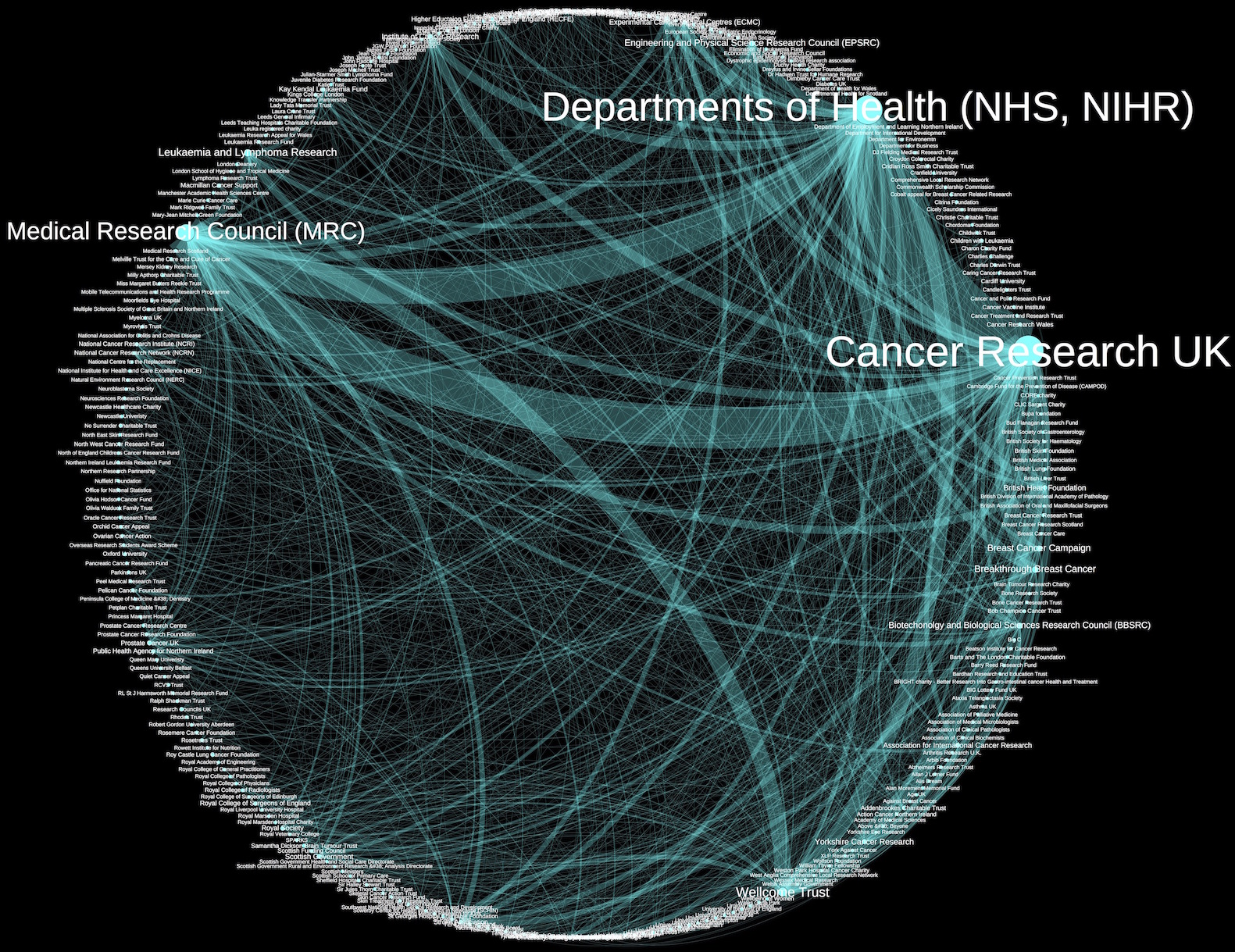}
\centering
\caption{Co-occurrence of UK funders in publication acknowledgement sections: nodes represent UK funding organisations (296) acknowledged in at least in two publications (the size of a node is proportional to the number of publications that acknowledged the funder the node represents, while the width of a tie is proportional to the number of times the two funders connected by the tie were jointly acknowledged in publications). The high-definition version of this figure is available at \url{https://dx.doi.org/10.6084/m9.figshare.2064897.v1}.
\newline\textit{Source: Authors' elaboration.}}
\label{fig:cofunding}
\end{figure}

Results are depicted in \Figref{cofunding}, where each node is a UK funder and the width of the tie between two nodes indicates the extent to which the two funders are jointly acknowledged in publications. For the sake of clarity in the visualisation, we reported UK funders that were acknowledged in at least two publications in our sample (296 funders). The analysis revealed that the UK funders most frequently jointly acknowledged in cancer research publications are the MRC, Cancer Research UK, and the UK Departments of Health (including the NHS in England and the devolved regions and NIHR). The majority of funders shown in  \Figref{cofunding} are external to the research host organisation undertaking the research. Nonetheless, it is worth noting that a small proportion of funders are research host organisations that also appear to support cancer research through internal funds.\footnote{Although some authors did acknowledge such contributions, generally, as discussed, authors did not acknowledge their employers as a funder. However, it is not possible to exclude these where they do occur. } \Tabref{co-occurence} also reports the co-occurrence matrix for the top-10 UK funders.

\setlength{\tabcolsep}{6pt}
\renewcommand{\arraystretch}{1.0}
\begin{table}\footnotesize
	\caption{\label{tab:co-occurence}Co-occurrence matrix (joint acknowledgements) of the top-10 UK funders.}
	\centering
{\begin{tabular}{lcccccccccc}
\hline\hline
\textbf{UK funder}         		& \textbf{1}   & \textbf{2}   & \textbf{3}  & \textbf{4}  & \textbf{5} & \textbf{6} & \textbf{7} & \textbf{8}  & \textbf{9} & \textbf{10} \\
\hline
1.\ Cancer Research UK                                  			& -   &     &    &    &   &   &   &    &   &    \\
2.\ Departments of Health (NHS, NIHR)                   		& 315 & -   &    &    &   &   &   &    &   &    \\
3.\ Medical Research Council                            			& 216 & 183 & -  &    &   &   &   &    &   &    \\
4.\ Wellcome Trust                                      				& 83  & 75  & 86 & -  &   &   &   &    &   &    \\
5.\ Leukaemia and Lymphoma Research                    		& 42  & 24  & 41 & 9  & - &   &   &    &   &    \\
6.\ Breakthrough Breast Cancer                          			& 51  & 52  & 15 & 10 & 3 & - &   &    &   &    \\
7.\ Breast Cancer Campaign                              			& 54  & 19  & 7  & 3  & 0 & 7 & - &    &   &    \\
8.\ Engineering and Physical Science Research Council   	& 44  & 46  & 43 & 7  & 2 & 4 & 2 & -  &   &    \\
9.\ Biotechnology and Biology Sciences Research Council & 21  & 10  & 25 & 15 & 9 & 1 & 1 & 15 & - &    \\
10.\ Yorkshire Cancer Research                          		& 19  & 14  & 8  & 1  & 3 & 4 & 9 & 0  & 5 & - \\

\hline\hline
\multicolumn{11}{l}{\footnotesize \textit{Source: Authors' elaboration.}}
\end{tabular}
}
\end{table}

\subsubsection{Funders' research portfolios}
The MeSH classification enables us to link funders acknowledged in publications and the content of publications in terms of medical topics. We focused on the third level of the MeSH classification and specifically on the `children' of ``Neoplasms by Site'' and ``Neoplasms by Histologic Type'' descriptors, which hold 68\% and 43\% of the publications in our sample, respectively. Overall, these descriptors classify 6,174 publications (82.2\%).\footnote{The publication count includes also descriptors at lower levels of the MeSH classification. 1,126 publications (15\%) are classified as ``Neoplasms'' at the first level of the MeSH tree, i.e.\ no additional MeSH descriptors at second or lower levels are reported. Other second-level descriptors under the ``Neoplasms'' are assigned to 934 publications.} The selected descriptors with the associated number of publications in our sample are reported in \Tabref{mesh}. The three areas with most publications are glandular and epithelial, digestive system, and urogenital neoplasms.

\setlength{\tabcolsep}{5pt}
\renewcommand{\arraystretch}{1.0}
\begin{table}[h]\footnotesize
	\caption{\label{tab:mesh}Cancer research areas.}
	\centering
{\begin{tabular}{lllc}
\hline\hline
\textbf{Descriptor}                            & \textbf{Abbreviation}                     & \textbf{Tree number} & \textbf{Number of} \\
\textbf{ }                            & \textbf{ }                     & \textbf{} & \textbf{publications} \\

\hline
\textbf{Neoplasms by Site}                     & -                                & \textbf{C04.588}     & \textbf{5,137}                 \\
\addtolength{\leftskip}{1em} Digestive System Neoplasms            & Digestive System                 & C04.588.274 & 1,253                  \\
\addtolength{\leftskip}{1em} Urogenital Neoplasms                  & Urogenital                       & C04.588.945 & 1,017                  \\
\addtolength{\leftskip}{1em} Breast Neoplasms                      & Breast                           & C04.588.180 & 1,008                  \\
\addtolength{\leftskip}{1em} Head and Neck Neoplasms               & Head \& Neck                     & C04.588.443 & 663                    \\
\addtolength{\leftskip}{1em} Endocrine Gland Neoplasms             & Endocrine Gland                  & C04.588.322 & 565                    \\
\addtolength{\leftskip}{1em} Thoracic Neoplasms                    & Thoracic                         & C04.588.894 & 533                    \\
\addtolength{\leftskip}{1em} Nervous System Neoplasms              & Nervous System                   & C04.588.614 & 373                    \\
\addtolength{\leftskip}{1em} Skin Neoplasms                        & Skin                             & C04.588.805 & 287                    \\
\addtolength{\leftskip}{1em} Bone Neoplasms                        & Bone                             & C04.588.149 & 236                    \\
\addtolength{\leftskip}{1em} Soft Tissue Neoplasms                 & Soft Tissue                      & C04.588.839 & 62                     \\
\addtolength{\leftskip}{1em} Eye Neoplasms                         & Eye                              & C04.588.364 & 61                     \\
\addtolength{\leftskip}{1em} Hematologic Neoplasms                 & Hematologic                      & C04.588.448 & 60                     \\
\addtolength{\leftskip}{1em} Abdominal Neoplasms                   & Abdominal                        & C04.588.33  & 46                     \\
\addtolength{\leftskip}{1em} Mammary Neoplasms, Animal             & Mammary (Animal)                 & C04.588.531 & 23                     \\
\addtolength{\leftskip}{1em} Pelvic Neoplasms                      & Pelvic                           & C04.588.699 & 13                     \\
\addtolength{\leftskip}{1em} Splenic Neoplasms                     & Splenic                          & C04.588.842 & 5                      \\
\addtolength{\leftskip}{1em} Anal Gland Neoplasms                  & Anal Gland                       & C04.588.83  & 1                      \\
                                      &                                  &             &                        \\
\textbf{Neoplasms by Histologic Type}          & -                                & \textbf{C04.557}     & \textbf{3,252}                 \\
\addtolength{\leftskip}{1em} Neoplasms, Glandular and Epithelial   & Glandular \& Epithelial          & C04.557.470 & 1,633                  \\
\addtolength{\leftskip}{1em} Neoplasms, Germ Cell and Embryonal    & Germ Cell \& Embryonal           & C04.557.465 & 652                    \\
\addtolength{\leftskip}{1em} Neoplasms, Nerve Tissue               & Nerve Tissue                     & C04.557.580 & 614                    \\
\addtolength{\leftskip}{1em} Leukemia                              & Leukemia                         & C04.557.337 & 435                    \\
\addtolength{\leftskip}{1em} Neoplasms, Connective and Soft Tissue & Connective \& Soft Tissue        & C04.557.450 & 328                    \\
\addtolength{\leftskip}{1em} Lymphoma                              & Lymphoma                         & C04.557.386 & 302                    \\
\addtolength{\leftskip}{1em} Nevi and Melanomas                    & Nevi \& Melanomas                & C04.557.665 & 229                    \\
\addtolength{\leftskip}{1em} Neoplasms, Plasma Cell                & Plasma Cell                      & C04.557.595 & 132                    \\
\addtolength{\leftskip}{1em} Neoplasms, Vascular Tissue            & Vascular Tissue                  & C04.557.645 & 115                    \\
\addtolength{\leftskip}{1em} Neoplasms, Complex and Mixed          & Complex \& Mixed                 & C04.557.435 & 62                     \\
\addtolength{\leftskip}{1em} Neoplasms, Gonadal Tissue             & Gonadal Tissue                   & C04.557.475 & 11                     \\
\addtolength{\leftskip}{1em} Lymphatic Vessel Tumors               & Lymphatic Vessel Tumors          & C04.557.375 & 10                     \\
\addtolength{\leftskip}{1em} Histiocytic Disorders, Malignant      & Histiocytic Disorders, Malignant & C04.557.227 & 5                      \\
\addtolength{\leftskip}{1em} Odontogenic Tumors                    & Odontogenic Tumors               & C04.557.695 & 1                 \\

\hline\hline
\multicolumn{4}{l}{\footnotesize \textit{Source: Authors' elaboration on the basis of MEDLINE/PubMed data.}}
\end{tabular}
}
\end{table}

\begin{figure}
\includegraphics[width=13cm]{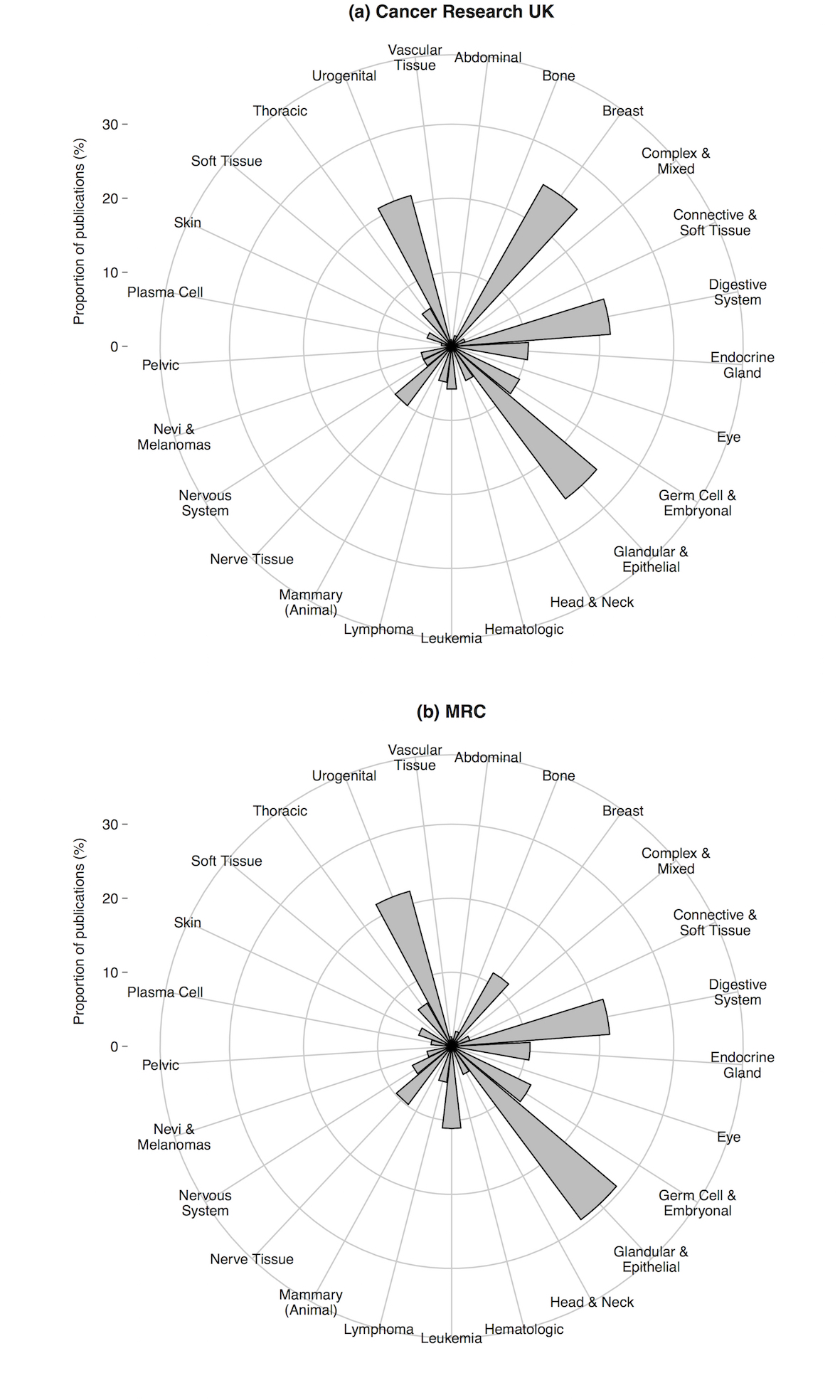}
\centering
\caption{Profile of four UK funders' research portfolios by cancer types (as based on the `abbreviated' MeSH descriptors assigned to publications): Cancer Research UK (a) and MRC (b).
\newline\textit{Source: Authors' elaboration.}}
\label{fig:profile1}
\end{figure}

\begin{figure}
\includegraphics[width=13cm]{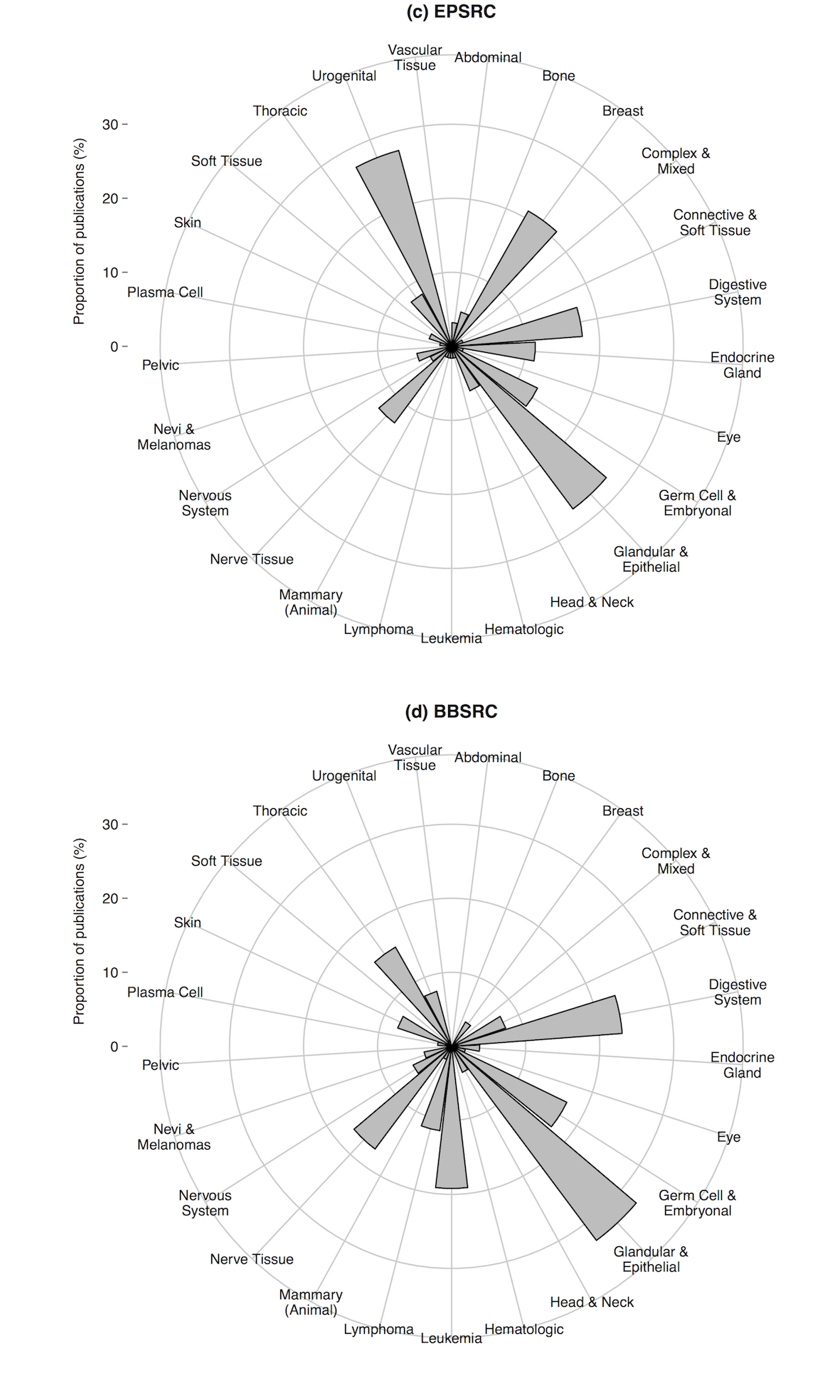}
\centering
\caption{Profile of four UK funders' research portfolios by cancer types (as based on the `abbreviated' MeSH descriptors assigned to publications): EPSRC (a) and BBSRC (b).
\newline\textit{Source: Authors' elaboration.}}
\label{fig:profile2}
\end{figure}

We used this classification to profile cancer research supported by four UK funders: Cancer Research UK, the Medical Research Council (MRC), the Engineering and Physical Science Research Council (EPSRC), and the Biotechnology and Biology Sciences Research Council (BBRSC) --- a broader set of cases is available elsewhere \citep{CRUK2014}. These funders were acknowledged in 24.2\%, 12.2\%, 2.4\%, and 2.3\% of the sub-sample of publications reporting funding data, respectively --- the low percentage for the last two cases is expected given that cancer research is not the main focus of these funders (indeed their strategic priorities extend beyond healthcare). \Figref{profile1, profile2} profiles the four funders in terms of the proportion of research that they supported in a given cancer domain, relative to overall number of publications that they supported in cancer.

\Figref{profile1}a shows that Cancer Research UK's support is mostly focused on breast, digestive system, glandular and epithelial, and urogenital areas. A similar profile can be observed for the case of MRC (\Figref{profile1}b) except for breast neoplasm, where CR-UK has a much more pronounced focus in its portfolio than MRC. EPSRC (\Figref{profile2}a) has also a similar portfolio in its cancer-related funded research (despite cancer not being the main focus of this funder). In the case of BBSRC (\Figref{profile2}b), the analysis shows that this funder tends to support research on germ cell and embryonal, thoracic, and nerve tissue neoplasms as well as on leukaemia and lymphoma. It is worth noting that about 15\% of the BBSRC-funded publications are within the thoracic neoplasm area, while less than 10\% of the publications supported by Cancer Research UK, EPSRC, and MRC fall within this area. Thoracic neoplasms include lung neoplasms, which account for the biggest cancer burden \citep{Stewart2014}.

\section{Discussion and conclusions}
At the present time there is an increasing pressure on research funding. At the same time, there is increasing access to data, facilitated by the Internet and improvements in computing power. These trends have combined to give new impetus for the use of funding data to analyse relationships between funding inputs and research outputs, such as publications. Funders have started to invest in large data infrastructures. For example, the US NIH's grants have been linked to their outcomes (publications and patents) in the RePORTER database, which is also made publicly available. Similarly, an increasing number of UK and international funders are relying on Researchfish to capture outcomes generated from their funded research, while the UK Clinical Research Collaboration has focused on the collection and analysis of data on the portfolios of scores of charitable and public sector funders. However, these `top-down' initiatives allow analysis only of research that funders have shared. Many funders that also contribute to funding systems are excluded. The analysis of funding data included in the acknowledgment \textit{paratext} of scientific publications this paper has focused is instead a `bottom-up' approach, which, in turn, can provide a more comprehensive perspective on the variety of funders involved in funding systems. This does not rely on a priori knowledge of the population of funders nor does it require funders' support to access the data. Nevertheless, significant challenges exist in the collection process (extraction and coding) of these data.

This paper has addressed a series of questions related to the use of this `bottom-up' approach to gathering funding data. To do so, we focused on a sample of 7,510 publications published in 2011 in cancer research by UK authors. First, the paper sets out a number of heuristics for guiding the collection and coding of funding data included in acknowledgements \textit{paratext}. These can inform future studies that aim to build datasets of funding data from publications as well as to develop natural language algorithms capable of extracting funding information from the acknowledgements \textit{paratext} automatically.

Second, we have examined the extent to which authors acknowledge funders in their publications. We specifically reviewed extant literature on the use funding data. These studies suggest that the reporting of funding data in publications is relatively limited in certain domains, such as some social sciences, due to their lower reliance on grant funding than the natural sciences, in which authors tend to acknowledge external funders more frequently. On the basis of reading and interpreting acknowledgements sections and surveying UK (corresponding) authors of a sample of publications that did not contain acknowledgements to funders, we estimated that in less than 3\% of the cases funding acknowledgements were entirely omitted when at least one external funder should have been disclosed.

Third, we provide evidence on the `quality' of funding data available in existing publication databases. Specifically, we compare MEDLINE/PubMed and ISI Web of Science (WoS), with those data that we manually collected and extracted following the heuristics we set out. On the one hand, we found that MEDLINE/PubMed has a very low \textit{recall} (about 42\%), thus many publications that include funding information are not identified as such by this database. Also, given the focus of this databases on few funders, those MEDLINE/PubMed records including funding information correctly listed funders in only about 25\% of the cases. On the other hand, WoS reported values of \textit{recall} and \textit{precision} well above 90\%, but the set of funders acknowledged in a publications were not correctly listed in about 32\% of the cases (i.e.\ the list of funders provided by the ``FU'' field could not be reconciled with the authors' reading of the acknowledgements \textit{paratext}). 

Fourth, through the application of bibliometric analysis to these data, we illustrated how funding data can be used to inform policy making on research funding. We used the data to provide a number of illustrative examples of major (charitable, governmental and private) funders supporting cancer research in the UK, showing the extent to which funders are co-funding research publications (i.e.\ being jointly acknowledged in the same publications), and profiling the research portfolios of these funders by cancer type.

Implications for development and use of funding data can be derived on the basis of these findings. Our review of extant literature on the proportion of publications with funding acknowledgements found a lack of studies on similar fields and regions over time. As a result, it remains difficult to determine whether the propensity to acknowledge funders has increased or not in recent years --- although one could argue that funders' efforts to link their funding to outputs and journals' efforts to boost transparency should yield an improvement, particularly in the biomedical field. 

Funders, journal editors and publishers have an important role to play in ensuring better reporting and transparency. Standardisation of the reporting modes of funding (e.g.\ consistent linking of particular funders to authors) would be a major contribution to the utility of funding data. It is often not possible to identify how much funding is linked to a given author or even to identify which author is acknowledging a funder linked to a publication. Also, authors can strategically (or mistakenly) acknowledge funders that have not supported the specific publications in which they are mentioned. It is generally not possible to link publications with the \textit{quanta} of research funding using funding acknowledgements \textit{paratext}. 

It is desirable that the contributions of all kinds of funders should be recognised. There is a clear tendency to acknowledge external funders, but not to acknowledge funding support from authors' employers and  funders that provide block funding (e.g. the UK Higher Education Funding Council was acknowledged in about 0.4\% of the publications included in our sample). Such omissions are a substantial hindrance to the accurate description of funding landscapes. Improvements in transparency and completeness of funding data would be to the benefit of analysts of funding systems, as the potential for comprehensiveness of the bottom-up approach are attractive.

The findings reported here also have important implications for future research on funding data. We provided evidence that the ``FU'' field (list of funders included in publications' acknowledgements) of WoS has some major limitations, with around one third of records in our sample not being correctly coded. These limitations may, however, be less prominent in areas of science that have lower funding intensity, for example with fewer funders per publication. It is worth noting that, the ``FX'' field (including the full-text of publications' acknowledgements) provided by WoS can be a helpful starting point to build datasets matching funding sources with scientific publications. Nonetheless, these data also require substantial cleaning and aggregation to ensure that all of the publications related to a given funder are appropriately linked. MEDLINE/PubMed can provide indications of funding activity for only a limited number of major US and non-US funders and therefore is not a suitable starting point for analysis of funding landscapes owing to the large number of funders that are not captured in the data at present.

Much of this paper has focused on cautions of the limitations of funding data, and naturally the analysis provided here suffers from many of the shortcomings we have identified, particularly in relation to data coverage. Those working with data outside biomedical sciences will likely find these problems too. Those seeking to provide longitudinal analysis on the dynamics of funding systems, which we have not explored here, will likely find the difficulties in applying our approach will multiply. Yet such efforts are worthwhile as bottom-up efforts to study funding systems have much to offer in terms of strategic intelligence for policy-making \citep{Rotolo2016}. This includes intelligence on active funders, joint-funding (as in the extent to which funders are jointly acknowledged in publications), and profiling of funders' research portfolios (e.g.\ which areas are more intensively supported by certain funders and where funders' interests overlap). Dynamic data can show how interaction between types of funder may be causally linked to certain benefits for particular types of research. This can aid the design of funding programmes and systems as well as improving our understanding of how funders may be complementary or interdependent.

\section*{Acknowledgements}
All the authors acknowledge the support of Cancer Research UK for the research project ``Exploring the Interdependencies of Research Funders in the UK''. MH and DR also acknowledge support from the UK Economic and Social Research Council for the award ``Mapping the Dynamics of Emergent Technologies'' (RES-360-25-0076) during which approaches used for this study were developed. DR further acknowledges the support of the People Programme (Marie Curie Actions) of the European Union's Seventh Framework Programme (FP7/2007-2013) (award PIOF-GA-2012-331107 - \href{http://www.danielerotolo.com/netgenesis}{"{\color{blue}NET-GENESIS: Network Micro-Dynamics in Emerging Technologies}}"). During the course of this research MH has received funding from the Higher Education Funding Council for England QR funding stream (as distributed through the UK's REF funding allocation system). All authors thank the University of Sussex for bearing residual costs of this research not met by the above named funders. The findings and observations contained in this paper are those of the authors and do not necessarily reflect the funders' views. We are thankful to the research assistance provided by Philippa Crane, Christopher Farrell, Abigail Mawer, Chelsea Pateman, and Tammy-Ann Sharp. We also are grateful to Aoife Regan, Emma Greenwood, Daniel Bridge, Jon Sussex, Ismael Rafols, Ben Martin, Paul Nightingale, Richard Sullivan, Virginia Acha, Michael O'Neill, Kevin Dolby, Shemila Nebhrajani, the two anonymous referees of the SPRU Working Paper Series (SWPS), and the two anonymous referees of the Journal of the Association for Information Science and Technology for their comments, criticisms and suggestions. We are also grateful for helpful feedback from audiences at the 2014 Eu-SPRI conference, the 2014 Global TechMining Conference, the 2015 Atlanta Conference, the 2015 Manchester Data Forum, and the 2015 British Council Newton Workshop on ``Science, Technology and Innovation in Neglected Diseases: Policies, Funding and Knowledge Creation'' in Belo Horizonte (Brazil). All remaining errors in the paper are those of the authors. We offer our kind regards to anyone using this piece of acknowledgement \textit{paratext} for analytical purposes.


\newpage
\singlespace
\bibliographystyle{apalike}
\bibliography{/Users/danielerotolo/Dropbox/References/bibtex_references/library.bib}

\end{document}